\documentclass[aps,prl,twocolumn,superscriptaddress,longbibliography]{revtex4-1}
\pdfoutput=1
\usepackage{graphicx}
\usepackage{amssymb,amsmath}
\usepackage{bm}
\usepackage{dcolumn}
\usepackage{subfigure}
\usepackage{float}
\usepackage[OT1]{fontenc} 
\usepackage{url}
\usepackage{mathrsfs}
\usepackage{slashed}
\usepackage{color}
\usepackage{verbatim}
\usepackage{enumitem}
\usepackage{txfonts}
\usepackage{soul,physics}
\usepackage[driverfallback=dvipdfm]{hyperref}
\hypersetup{pdfpagemode=FullScreen,colorlinks=true,breaklinks,urlcolor=blue,linkcolor=blue,citecolor=blue}
\usepackage{natbib}
\usepackage{environ}

\usepackage{subfigure}
\usepackage{comment}
\usepackage{hyperref}
\usepackage{amssymb}
\usepackage{bm}
\usepackage{graphicx}
\usepackage{amsmath,amssymb}

\usepackage{pdfpages} % include pdfs
\usepackage{pgffor} % for loops

\makeatletter
\AtBeginDocument{\let\LS@rot\@undefined}
\makeatother

\def\supplementfilename{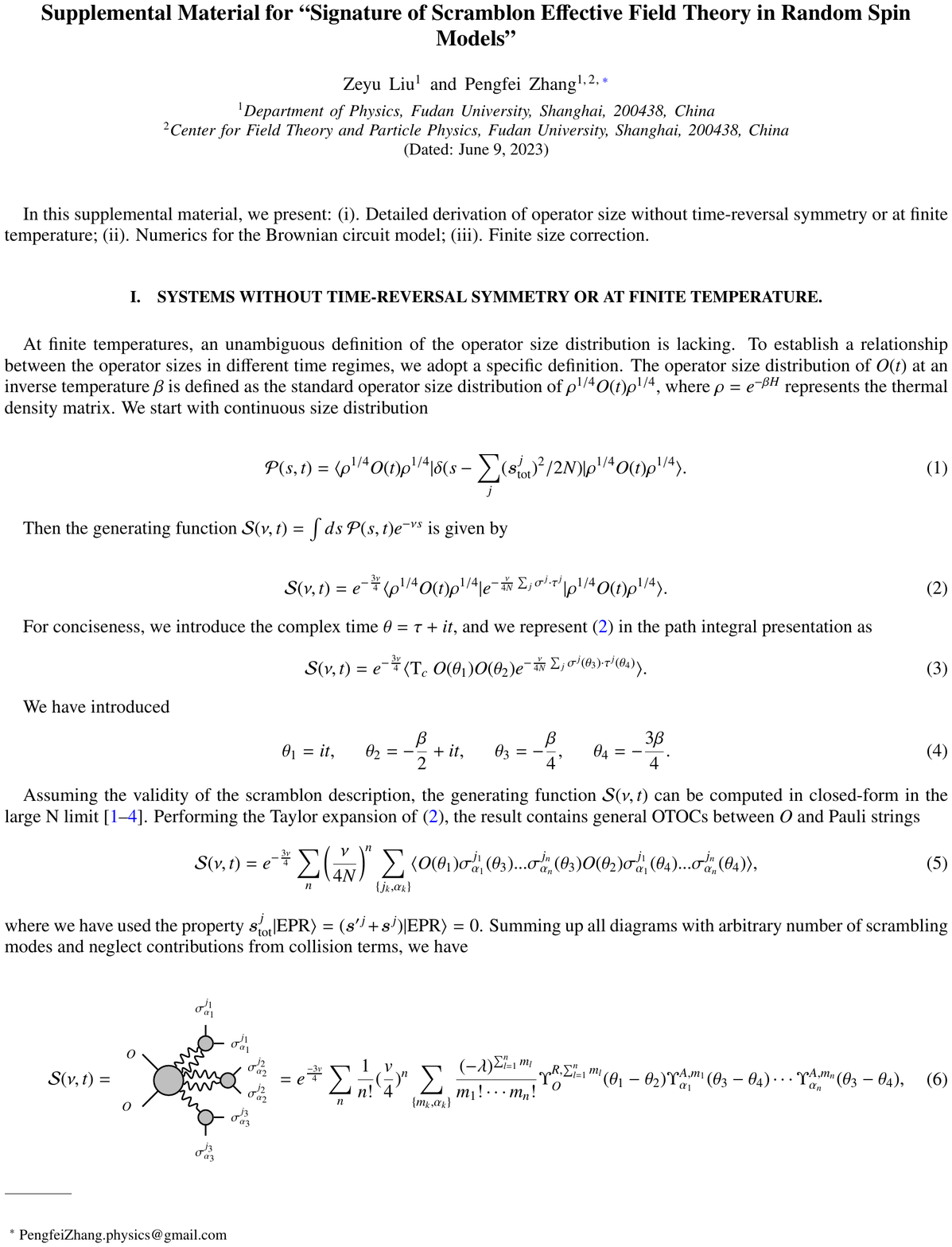}

\pdfximage{\supplementfilename}
\def\numbersupplementpages{\the\pdflastximagepages}

\newif\ifarXiv
\arXivtrue 
\usepackage{tikz,fp}
\usepackage{tikz-cd}
\usetikzlibrary{arrows}
\usetikzlibrary{intersections}
\usetikzlibrary{shapes.geometric}
\usetikzlibrary{decorations.pathmorphing, patterns,shapes,fixedpointarithmetic}
\usetikzlibrary{decorations.markings}

\pgfdeclarepatternformonly{south west lines}{\pgfqpoint{-0pt}{-0pt}}{\pgfqpoint{3pt}{3pt}}{\pgfqpoint{3pt}{3pt}}{
	\pgfsetlinewidth{0.4pt}
	\pgfpathmoveto{\pgfqpoint{0pt}{0pt}}
	\pgfpathlineto{\pgfqpoint{3pt}{3pt}}
	\pgfpathmoveto{\pgfqpoint{2.8pt}{-.2pt}}
	\pgfpathlineto{\pgfqpoint{3.2pt}{.2pt}}
	\pgfpathmoveto{\pgfqpoint{-.2pt}{2.8pt}}
	\pgfpathlineto{\pgfqpoint{.2pt}{3.2pt}}
	\pgfusepath{stroke}}

\tikzset{
	mid arrow/.style={postaction={decorate,decoration={
				markings,
				mark=at position .575 with {\arrow{stealth}}
	}}},
	near arrow/.style={postaction={decorate,decoration={
				markings,
				mark=at position .275 with {\arrow{stealth}}
	}}},
	far arrow/.style={postaction={decorate,decoration={
				markings,
				mark=at position .800 with {\arrow{stealth}}
	}}},
	snake arrow/.style={fixed point arithmetic, decorate, decoration={snake,amplitude=2pt, segment length=11pt},postaction={decoration={markings,mark=at position 0.625 with {\arrow{stealth}}},decorate}},
}
\tikzset{
  baseline = -0.5ex,
  wavy/.style = {
    thick,
    decorate,
    decoration={snake,amplitude=2pt,segment length=5pt}},
  sdot/.style = {
    circle,
    draw=none,
    fill=black,
    minimum size=2.5pt,
    inner sep=0pt},
  bdot/.style = {
    circle,
    draw=none,
    fill=black,
    minimum size=4pt,
    inner sep=0pt},
  svertex/.style = {
    circle,
    draw=black,
    thick,
    fill=lightgray,
    minimum size=8pt,
    inner sep=1pt},
  bvertex/.style = {
    circle,
    draw=black,
    thick,
    fill=lightgray,
    minimum size=24pt},
  bvertexsmall/.style = {
    circle,
    draw=black,
    thick,
    fill=lightgray,
    minimum size=7pt},
  bvertexnormal/.style = {
    circle,
    draw=black,
    thick,
    fill=lightgray,
    minimum size=16pt},
  dvertex/.style = {
    circle,
    draw=black,
    thick,
    fill=gray,
    minimum size=25pt}}

\NewEnviron{myequation}{%
    \begin{equation}
    \scalebox{0.91}{$\displaystyle{\BODY}$}
    \end{equation}
    }

\begin{document}
	
	\title{Signature of Scramblon Effective Field Theory in Random Spin Models }
	
	\author{Zeyu Liu}
	\affiliation{Department of Physics, Fudan University, Shanghai, 200438, China}
	\author{Pengfei Zhang}
	\thanks{PengfeiZhang.physics@gmail.com}
	\affiliation{Department of Physics, Fudan University, Shanghai, 200438, China}
    \affiliation{Center for Field Theory and Particle Physics, Fudan University, Shanghai, 200438, China}
	\date{\today}
	\begin{abstract}
    Information scrambling refers to the propagation of information throughout a quantum system. Its study not only contributes to our understanding of thermalization but also has wide implications in quantum information and black hole physics. Recent studies suggest that information scrambling is mediated by collective modes called scramblons. However, a criterion for the validity of scramblon theory in a specific model is still missing. In this work, we address this issue by investigating the signature of the scramblon effective theory in random spin models with all-to-all interactions. We demonstrate that, in scenarios where the scramblon description holds, the late-time operator size distribution can be predicted from its early-time value, requiring no free parameters. As an illustration, we examine whether Brownian circuits exhibit a scramblon description and obtain a positive confirmation both analytically and numerically. We also discuss the prediction of multiple-quantum coherence when the scramblon description is valid. Our findings provide a concrete experimental framework for unraveling the scramblon field theory in random spin models using quantum simulators.
	\end{abstract}
	
	\maketitle

	\emph{ \color{blue!60}Introduction.--}
    Quantum thermalization \cite{PhysRevE.50.888,PhysRevA.43.2046} in many-body systems requires the initial local information to become scrambled throughout the entire system under unitary evolutions \cite{Hayden:2007cs,Sekino:2008he,Shenker:2014cwa}. This process, known as information scrambling,  serves as a crucial link connecting condensed matter, quantum information, and gravity physics. In the Heisenberg picture, information scrambling is described by the growth of operator size \cite{Roberts:2014isa,Nahum:2017yvy,vonKeyserlingk:2017dyr,Khemani:2017nda,Hunter-Jones:2018otn,Chen:2019klo,PhysRevLett.122.216601,Chen:2020bmq, Lucas:2020pgj,Yin:2020oze,Zhou:2021syv,Dias:2021ncd,2021PhRvR...3c2057W,Zhang:2022knu,Zhang:2022fma,Jian:2020qpp}, with its expectation value being associated with out-of-time-order correlators (OTOCs). In $N$-qubit chaotic systems with all-to-all interactions, the average operator size exhibits universal exponential growth in the early-time regime $t\ll t_{\text{sc}}$, serving as a signature of quantum chaos. Here $t_{\text{sc}}\propto \ln N$ denotes the scrambling time. Conversely, understanding information scrambling in the late-time regime with $t\sim t_{\text{sc}}$ poses greater challenges due to its non-perturbative nature.

    Recently, inspired by the diagrammatic analysis in the Sachdev-Ye-Kitaev (SYK) model \cite{kitaev2015simple,Maldacena:2016hyu,Kitaev:2017awl,RevModPhys.94.035004}, the scramblon effective field theory (SEFT) was proposed as a unified description of information scrambling in both regimes \cite{Gu:2021xaj}. The key assumption is that for long time separations, only out-of-time-order correlations are essential, which are mediated by collective modes called scramblons \cite{Gu:2018jsv,Gu:2021xaj}. The SEFT is specified by a set of Feynman rules. For systems with time-reversal symmetry, this includes the scramblon propagator $\lambda = e^{\varkappa t}/C$ (with quantum Lyapunov exponent $\varkappa$ and numerical factor $C\propto N$) and the scattering vertex between a pair of Hermitian operators $\hat{O}$ and $m$ scramblons
    \begin{equation}\label{eqn:h}
    \begin{tikzpicture}
    \node[bvertexnormal] (R) at (-30pt,0pt) {};
    \draw[thick] (R) -- ++(135:18pt) node[left]{\scriptsize$\hat{O}^\dagger(t_1)$};
    \draw[thick] (R) -- ++(-135:18pt) node[left]{\scriptsize$\hat{O}^{\ }(t_2)$};
    \draw[wavy] (R) -- ++(45:18pt) node[left]{};
    \draw[wavy] (R) -- ++(-45:18pt) node[left]{};
    \draw[wavy] (R) -- ++(0:18pt) node[left]{};
    \end{tikzpicture}=\Upsilon^{m}_{O }(t_{12})=\int_0^\infty dy~y^m h_{ O }(y,t_{12}).
    \end{equation}
    Here the wavy lines represent the scramblon modes. In the remaining part of the manuscript, we omit the argument $t_{12}$ when $t_{12}=0$. In particular, we have $\Upsilon^{m}_{O}=\langle\hat{O}^2\rangle$, where we define $\langle \hat{M} \rangle=\mathcal{D}^{-1}\text{tr}[\hat{M}]$ with Hilbert space dimension $\mathcal{D}$. 
    
    By employing the SEFT, analytical results for OTOCs \cite{Gu:2021xaj,Zhang:2023vpm}, operator size distribution \cite{Zhang:2022fma,Zhang:2022knu,newpaper2}, and quantum teleportation \cite{newpaper} can be obtained in SYK-like models. However, the exact class of models where the scramblon theory is valid remains largely unexplored. The main challenge lies in the absence of a valid diagrammatic expansion in general $0+1$-D models. On the other hand, significant advancements in the quantum simulation of many-body systems have been witnessed in recent years. In particular, considerable attention has been paid to the experimental study of information scrambling by measuring OTOCs \cite{2015Natur.528...77I,Li:2016xhw,Garttner:2016mqj,2019Sci...364..260B,2019arXiv190206628S,2019Natur.567...61L,2020PhRvL.124x0505J,Blok:2020may,2021PhRvA.104a2402D,2021PhRvA.104f2406D,Mi:2021gdf,Cotler:2022fin,2022PhRvA.105e2232S}. Furthermore, concrete protocols have been proposed for measuring the operator size distribution \cite{Qi:2019rpi}. 

\begin{figure}
    \centering
    \includegraphics[width=0.83\linewidth]{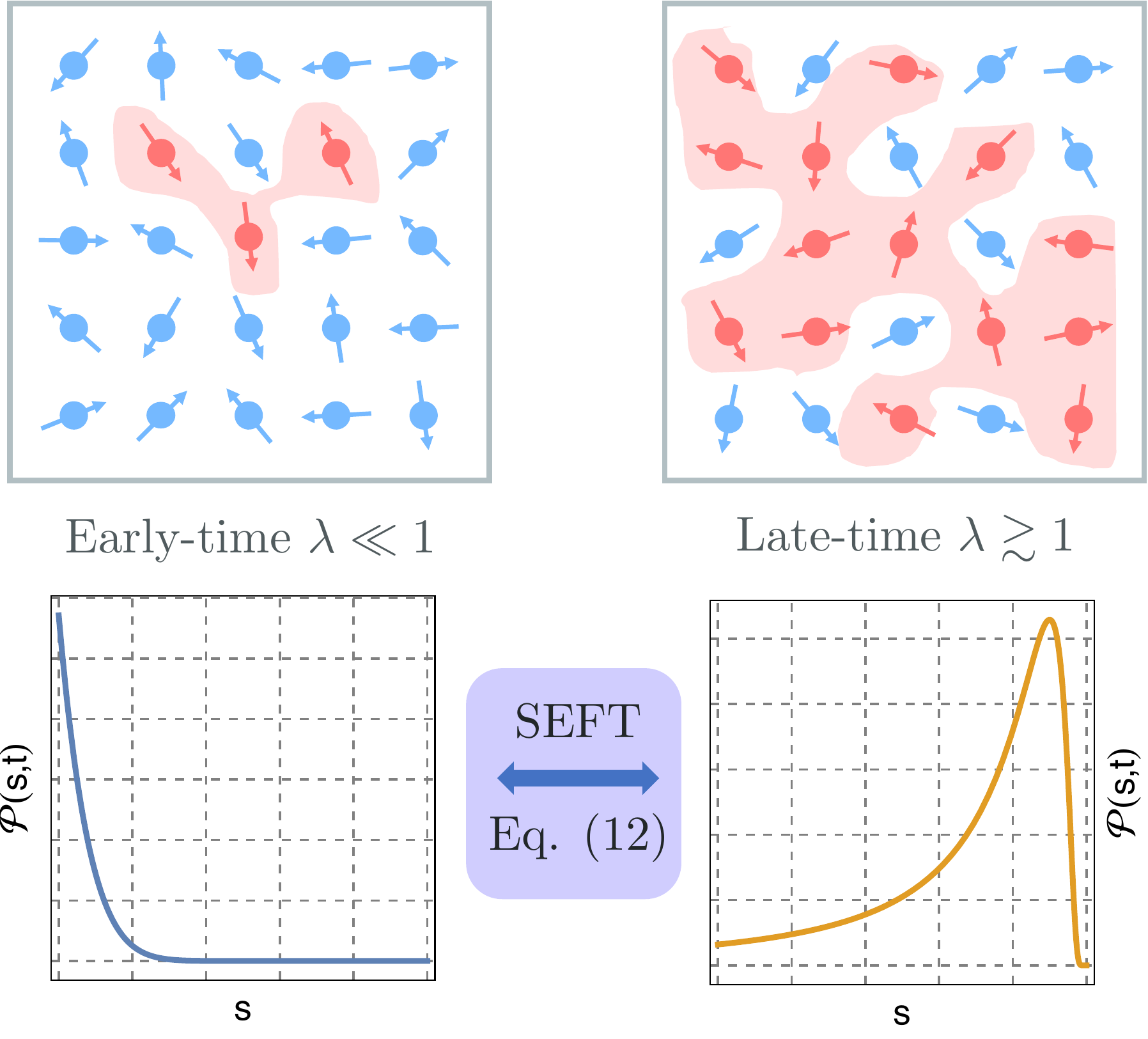} \label{fig1}
    \caption{The schematics illustrate the consistency relation between the early-time operator size distribution and the late-time operator size distribution, valid under the assumption of SEFT.}
\end{figure}
    In this work, motivated by these developments, we propose a scheme to test the validity of SEFT on quantum simulators. We focus on random spin models with all-to-all interactions, which has been realized in various experimental platforms. By generalizing the size-charge relation for Majorana systems, we derive the scramblon theory prediction of the operator size distribution in random spin models. The results show that the late-time operator size distribution can be predicted from its early-time value without any prior knowledge of $C$ or $\Upsilon^{m}_{ O }$. This consistency relation serves as a smoking gun for the SEFT. As an illustration, we apply our scheme to Brownian circuits \cite{Zhou:2018snw}, where the operator size distribution can be efficiently simulated by solving classical differential equations. Both analytical and numerical results confirm the validity of its SEFT description. We also discuss the prediction of multiple-quantum coherence (MQC) when the SEFT is valid. Our result is of fundamental interest for understanding the scrambling dynamics in generic quantum systems.
\vspace{5pt}

    \emph{ \color{blue!60}Operator size \& total spin.--} 
    We focus on systems of $N$ spins $\hat{\bm{s}}^j=\hat{\bm{\sigma}}^j/2$ with time-reversal-invariant all-to-all interactions. Here $j=$1, 2, ..., $N$ labels different spins. Protocols without the time-reversal symmetry are discussed in later. An Hermitian operator $\hat{O}(t)$ can be expanded using the complete basis of Pauli operators 
    \begin{equation}\label{eqn:operatorexpansion}
     \hat{O}(t)=\sum_{n} \sum_{j_{1}<\cdots <j_{n}} \sum_{\alpha_{1} \cdots \alpha_{n}} C_{j_{1}\alpha_{1} \cdots j_{n} \alpha_{n}}\hat \sigma_{\alpha_{1}}^{j_{1}}\hat \sigma_{\alpha_{2}}^{j_{2}} \cdots \hat\sigma_{\alpha_{n}}^{j_{n}} ,
    \end{equation}
    where $\alpha_{k}\in\{x, y, z\}$ labels different Pauli matrices. The length $n$ of $\sigma_{\alpha_{1}}^{j_{1}} \cdots \sigma_{\alpha_{n}}^{j_{n}} $ is defined as the size of this basis. The operator size distribution $P(n,t)$ counts the operator weight on bases with size $n$ as
    \begin{equation}
    P(n, t)=\sum_{j_{1}<\cdots <  j_{n}} \sum_{\alpha_{1} \cdots \alpha_{n}} (C_{j_{1} \alpha_{1} \cdots j_{n} \alpha_{n}})^{2}, 
    \end{equation}
    with $n\in\{1,2,...,N\}$. The total probability is then conserved $\sum_n P(n,t)=1$ due to the unitarity if we normalize $\langle\hat{O}^2\rangle=1$. In the thermodynamic limit $N\gg 1$, we further introduce a continuous variable $s=n/N \in[0,1]$ with continuous distribution $\mathcal{P}(s,t)\equiv NP(sN,t)$, which is normalized as $\int_0^1 ds ~\mathcal{P}(s,t)=1$. 

    %In this work, we will focus on the operator size distribution of single-site Pauli operators $\hat{O}=\hat{\sigma}^k_\alpha$. 

    In Ref. \cite{Qi:2018bje}, the authors establish a connection between the operator size in Majorana systems and the charge operator in the doubled system. This formulation enables a systematic investigation of the distribution of operator sizes within the framework of the SEFT. We present a generalization of this relation to spin models. We first introduce an auxiliary system consisting of $N$ spins $\hat{\bm{s}}'^j=\hat{\bm{\tau}}^j/2$. The doubled system is prepared in the tensor product of singlet states $|\text{EPR}\rangle=\prod_{\otimes j}|\text{S}\rangle_j=\prod_{\otimes j}(\ket{\uparrow}_{\sigma^j}\ket{\downarrow}_{\tau^j}-\ket{\downarrow}_{\sigma^j}\ket{\uparrow}_{\tau^j})$. Consequently, we have $\hat{\bm{s}}^j_{\text{tot}}|\text{EPR}\rangle=(\hat{\bm{s}}'^j+\hat{\bm{s}}^j)|\text{EPR}\rangle=0$. To study the size of an operator $\hat{O}(t)$, we map the operator to a state by $|O(t)\rangle=O(t)|\text{EPR}\rangle$ \cite{Qi:2018bje}. The Eq. \eqref{eqn:operatorexpansion} then becomes an expansion in orthonormal states. The crucial observation is 
    \begin{equation}
    |\text{T},\alpha\rangle_j\equiv\hat{\sigma}^j_\alpha|\text{S}\rangle_j,\ \ \ \ \ (\hat{\bm{s}}^j_{\text{tot}})^2|\text{T},\alpha\rangle_j=2|\text{T},\alpha\rangle_j.
    \end{equation}
    Here, $\text{T}$ indicates that the state is a spin-triplet. From a group theoretical perspective, this is due to $\hat{\sigma}^j_\alpha$ being a spin-$1$ spherical tensor operator. Consequently, the size of a string of Pauli operators is equal to the number of triplet states, counting over different site $j$. This gives 
    \begin{equation}\label{eqn:expPst}
    \mathcal{P}(s, t)=\langle \hat{O}(t)| \delta(s-\sum_j(\hat{\bm{s}}^j_{\text{tot}})^2/2N)|\hat{O}(t)\rangle.
    \end{equation}
    When calculating the operator size distribution, it is more convenient to introduce the generating function $\mathcal{S}(\nu,t)=\int ds~\mathcal{P}(s,t)e^{-\nu s}$. Using Eq. \eqref{eqn:expPst}, we find
    \begin{equation}\label{eqn:expSnut}
    \mathcal{S}(\nu, t)=e^{-\frac{3\nu}{4}}\langle \hat{O}(t)|e^{-\frac{\nu}{4N}\sum_j\bm{\sigma}^j\cdot \bm{\tau}^j}|\hat{O}(t)\rangle.
    \end{equation}

    Note that Eq. \eqref{eqn:expPst} and Eq. \eqref{eqn:expSnut} not only serve as a starting point for theoretical calculations but also provide a concrete protocol for measuring the operator size distribution on quantum simulators. This protocol requires the ability to prepare the initial EPR state and reverse the Hamiltonian, which are feasible with state-of-the-art cold atom systems and superconducting qubits.

\begin{figure}[t]\centering
\begin{tabular}{c@{\hspace{0.5cm}}c}
\begin{tikzpicture}[scale=1.0]
\filldraw[color=red!60, fill=red!5, dotted,thick](20pt,20pt) circle (8pt);
\node[bvertexnormal] (R) at (-0pt,0pt) {};
\node[svertex] (A1) at (20pt,20pt) {};
\node[svertex] (A2) at (32pt,0pt) {};
\node[svertex] (A3) at (20pt,-20pt) {};

\draw[thick] (R) -- ++(135:20pt) node[left]{\scriptsize$\hat{O}^\dagger$};
\draw[thick] (R) -- ++(-135:20pt) node[left]{\scriptsize$\hat{O}^{\ }$};
\draw[thick] (A1) -- ++(90:10pt) node[above]{\scriptsize$\hat{\sigma}^{j_1}_{\alpha_1}$};
\draw[thick] (A1) -- ++(0:10pt) node[right]{\scriptsize$\hat{\sigma}^{j_1}_{\alpha_1}$};
\draw[thick] (A3) -- ++(0:10pt) node[right]{\scriptsize$\hat{\sigma}^{j_3}_{\alpha_3}$};
\draw[thick] (A3) -- ++(-90:10pt) node[below]{\scriptsize $\hat{\sigma}^{j_3}_{\alpha_3}$};
\draw[thick] (A2) -- ++(45:10pt) node[right]{\scriptsize$\hat{\sigma}^{j_2}_{\alpha_2}$};
\draw[thick] (A2) -- ++(-45:10pt) node[right]{\scriptsize$\hat{\sigma}^{j_2}_{\alpha_2}$};

\draw[wavy] (R) to[out=60,in=210] (A1);
\draw[wavy] (R) to[out=30,in=240] (A1);
\draw[wavy] (R) to (A3);
\draw[wavy] (R) to[out=15,in=165] (A2);
\draw[wavy] (R) to[out=-15,in=195] (A2);
\end{tikzpicture}
&
\begin{tikzpicture}[scale=0.95]
\filldraw[color=red!60, fill=red!5, dotted,thick](0pt,0pt) circle (15pt);
\node[bvertexnormal] (R) at (-0pt,0pt) {};
\node[svertex] (A1) at (30pt,20pt) {};
\node[svertex] (A3) at (30pt,-20pt) {};

\draw[thick] (R) -- ++(135:20pt) node[left]{\scriptsize$\hat{O}(t)^\dagger$};
\draw[thick] (R) -- ++(-135:20pt) node[left]{\scriptsize$\hat{O}(t)^{\ }$};
\draw[thick] (A1) -- ++(70:10pt) node[above]{\scriptsize$\hat{\sigma}^{j_1}_{\alpha_1}$};
\draw[thick] (A1) -- ++(-20:10pt) node[right]{\scriptsize$\hat{\sigma}^{j_1}_{\alpha_1}$};
\draw[thick] (A3) -- ++(20:10pt) node[right]{\scriptsize$\hat{\sigma}^{j_2}_{\alpha_2}$};
\draw[thick] (A3) -- ++(-70:10pt) node[below]{\scriptsize $\hat{\sigma}^{j_2}_{\alpha_2}$};

\draw[wavy] (R) to (A1);
\draw[wavy] (R) to (A3);

\end{tikzpicture}
\vspace{3pt}\\
(a) General diagram for $\mathcal{S}(\nu,t)$& (b) The early-time limit $(\lambda \ll 1)$
\end{tabular}
\caption{Typical diagrams of $\mathcal{S}(\nu,t)$ are shown in (a) for the full time range and (b) in the early-time limit $\lambda \ll 1$. Wavy lines represent the propagator of scrambling modes, while solid lines represent the microscopic operators $\hat{O}(t)$ or $\hat{\sigma}^j_\alpha$. The vertices within the dashed red circle become identical for $\hat{O}=\hat{\sigma}^{j_1}_{\alpha_1}$, indicating a consistency relation between different time regimes. 
}
\label{fig:Diagram}
\end{figure}
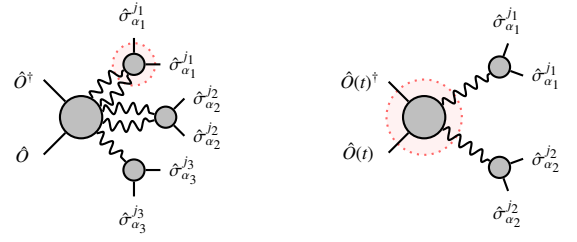

\vspace{5pt}

    \emph{ \color{blue!60}SEFT calculation.--} 
    Assuming the validity of the scramblon description, the generating function $\mathcal{S}(\nu, t)$ can be computed in closed-form. Performing the Taylor expansion of \eqref{eqn:expSnut}, the result contains general OTOCs between $\hat{O}$ and Pauli strings 
    \begin{equation}\label{eqn:expSnutTaylor}
    \mathcal{S}=e^{-\frac{3\nu}{4}}\sum_{m}\left(\frac{\nu}{4N}\right)^m\sum_{\{j_k,\alpha_k\}}\langle \hat{O}(t)\hat\sigma^{j_1}_{\alpha_1}...\hat\sigma^{j_m}_{\alpha_m}\hat{O}(t)\hat\sigma^{j_1}_{\alpha_1}...\hat\sigma^{j_m}_{\alpha_m}\rangle.
    \end{equation}
    We have neglected contributions from collision terms where $j_k=j_{k'}$, which become negligible in the thermodynamic limit $N\rightarrow \infty$. In the SEFT picture, each pair of $\sigma^{j_k}_{\alpha_k}$ operators can emit an arbitrary number of scramblons, all of which are ultimately absorbed by $\hat{O}$ operators in the future. A typical diagram with $m=3$ is shown in FIG. \ref{fig:Diagram} (a). By summing up all possible configurations, result can be expressed as 
    \begin{equation}
    \mathcal{S}=e^{-\frac{3\nu}{4}}\sum_{m}\frac{1}{m!}\left(\frac{\nu}{4}\right)^m\sum_{\{n_k,\alpha_k\}}\frac{(-\lambda)^{\sum_m{n_m}}}{m!}\Upsilon^{\sum_m{n_m}}_O \Upsilon_{\alpha_1}^{n_1}...\Upsilon_{\alpha_m}^{n_m}.
    \end{equation}
    Here, we define averaged vertex functions as $\Upsilon_\alpha^m \equiv \sum_j \Upsilon_{\sigma^j_\alpha}^m/N$ for conciseness. In comparison to the results for Majorana fermions \cite{Zhang:2022fma,Zhang:2022knu}, we need to include an additional summation over $\alpha$ to account for the contributions from different Pauli matrices. By utilizing Eq. \eqref{eqn:h}, we can perform the summation and obtain 
    \begin{equation}\label{eq:Snut}
    \mathcal{S}(\nu,t)=\int_0^\infty dy~h_O(y) e^{-\frac{\nu}{4}({3}-\sum_\alpha f_{\alpha}(\lambda y))},
    \end{equation}
    where we have $f_O(x)\equiv \sum_m(-x)^m \Upsilon_O^m/m!=\int dy~h_{ O }(y)e^{-xy}$ describing the perturbed two-point functions with scramblon fields \cite{Gu:2021xaj}. The operator size distribution can then be obtained by taking the inverse Laplace transform $\mathcal{P}(s,t)=\int_0^\infty dy~h_O(y)~ \delta\Big(s-\frac{{3}-\sum_\alpha f_{\alpha}(\lambda y)}{4}\Big).$
\vspace{5pt}

     \emph{ \color{blue!60}Consistency relation.--}    
     We are ready to show that Eq. \eqref{eq:Snut} indicates non-trivial consistency relation of the operator size distribution when the SEFT description is valid. We take an average of $\mathcal{\mathcal{P}}(s,t)$ over single Pauli operators $\hat O=\hat{\sigma}^j_\alpha$
     \begin{equation}\label{eq:averagePst}
     \overline{\mathcal{P}}(s,t)=\int_0^\infty dy~\overline{h}(y) \delta\Big(s-s_{\text{sc}}({{1}-\overline{f}(\lambda y))}\Big),
     \end{equation}
     with $\overline{g}(x)\equiv\sum_\alpha g_\alpha(x)/3\equiv\sum_{\alpha,j}g_{\sigma_\alpha^j}(x)/3N$. For systems in which different spins are equivalent, the averaging over $j$ can be omitted. $s_{\text{sc}}=3/4$ is the typical size for a maximally scrambled operator in spin models, where all possible operators appear with equal probability. This can be seen by taking $\lambda \rightarrow \infty$ in Eq. \eqref{eq:averagePst}, which gives $\overline{\mathcal{P}}(s,\infty)=\delta(s-s_{\text{sc}})$ for chaotic models with $\overline{f}(\infty)=0$ \cite{Gu:2021xaj}. It is worth noting that a similar expression holds for systems with Majorana fermions \cite{Zhang:2022fma}, with $s_{\text{sc}}=1/2$. Therefore, the consistency relation derived in this section can also be applied to Majorana systems.

     The key observation is that the averaged generating function at different times $\nu$ and $t$ relies solely on a single function $\overline{h}(y)$, as $\overline{f}(x)$ is its Laplace transform. Our strategy is to first express $\overline{h}(y)$ in terms of the early-time size distribution with $N^{-1}\ll \lambda \ll 1$, and then establish its relation to the distribution at late times. A diagrammatic illustration is presented in FIG. \ref{fig:Diagram}.  In the early-time regime with $N^{-1}\ll \lambda_0=e^{\varkappa t_0}/C \ll 1$, we expand $\overline{f}(\lambda_0 y)=1-\lambda_0 y \overline{\Upsilon}^{1}$. Using Eq. \eqref{eq:averagePst}, we find
     \begin{equation}
     \begin{aligned}
     \overline{\mathcal{P}}(s,t_0)&=({s_{\text{sc}}\lambda_0 \overline{\Upsilon}^{1}})^{-1}\overline{h}\Big(s({s_{\text{sc}}\lambda_0 \overline{\Upsilon}^{1}})^{-1}\Big),\\
     \overline{s_0}&=\int_0^1 ds~s~\overline{\mathcal{P}}(s,t_0)=s_{\text{sc}}\lambda_0 (\overline{\Upsilon}^{1})^2.
     \end{aligned}
     \end{equation}
     The main obstacle in extracting $\overline{h}(y)$ from $\overline{\mathcal{P}}(s,t_0)$ lies in the unknown coefficients $C$ and $\overline{\Upsilon}^1$. Fortunately, these coefficients ultimately cancel out, resulting in no free parameters in the consistency relation. To demonstrate this, let us consider the distribution \eqref{eq:averagePst} for a general time $\lambda$. By introducing $y=s_1 /({s_{\text{sc}}\lambda_0 \overline{\Upsilon}^{1}})$, it becomes straightforward to show that
     \begin{myequation}\label{eq:relation}
     \begin{aligned}
     %\overline{\mathcal{P}}(s,t)=&\int_{0}^{\infty} ds'~\overline{\mathcal{P}}(s',t_{0})\delta\left(s-s_{\text{sc}}+s_{\text{sc}}\overline{\mathcal{S}}\left(\frac{s'e^{\varkappa(t-t_0)}}{s_{\text{sc}}\overline{s_0}},t_0\right)\right).
     \overline{\mathcal{P}}(s,t)=&\int_{0}^{\infty} ds_1~\overline{\mathcal{P}}(s_{1},t_{0})\delta\bigg(s-s_{\text{sc}}\\&+s_{\text{sc}}\int_{0}^{\infty}\mathrm{d}s_{2}\overline{\mathcal{P}}(s_{2},t_{0})\text{exp}\Big(-\frac{s_{1}s_{2}} {s_{\text{sc}}\overline{s_{0}}}e^{\varkappa(t-t_{0})}\Big)\bigg).
     \end{aligned}
     \end{myequation}
      This is the main result for this work. Note that we can safely extend the integral to $[0,\infty]$ since the early-time distribution $\overline{\mathcal{P}}(s,t_{0})$ is only support near $s\ll 1$. Since \eqref{eq:relation} should be valid for arbitrary $s$ and $t$, it is highly restrictive and serves as a signature of SEFT in systems with all-to-all interactions. Eq. \eqref{eq:relation} can also be applied to predict the averaged operator size 
     \begin{equation}\label{eq:avgsize}
     \frac{\overline{s}}{s_{\text{sc}}}=1-\int_{0}^{\infty} \mathrm{d}s_{1} ds_{2}\overline{\mathcal{P}}(s_{1},t_{0})\overline{\mathcal{P}}(s_{2},t_{0})e^{-\frac{s_{1}s_{2}} {s_{\text{sc}}\overline{s_{0}}}e^{\varkappa(t-t_{0})}}.
    \end{equation}

\begin{figure}
    \centering
    \includegraphics[width=0.85\linewidth]{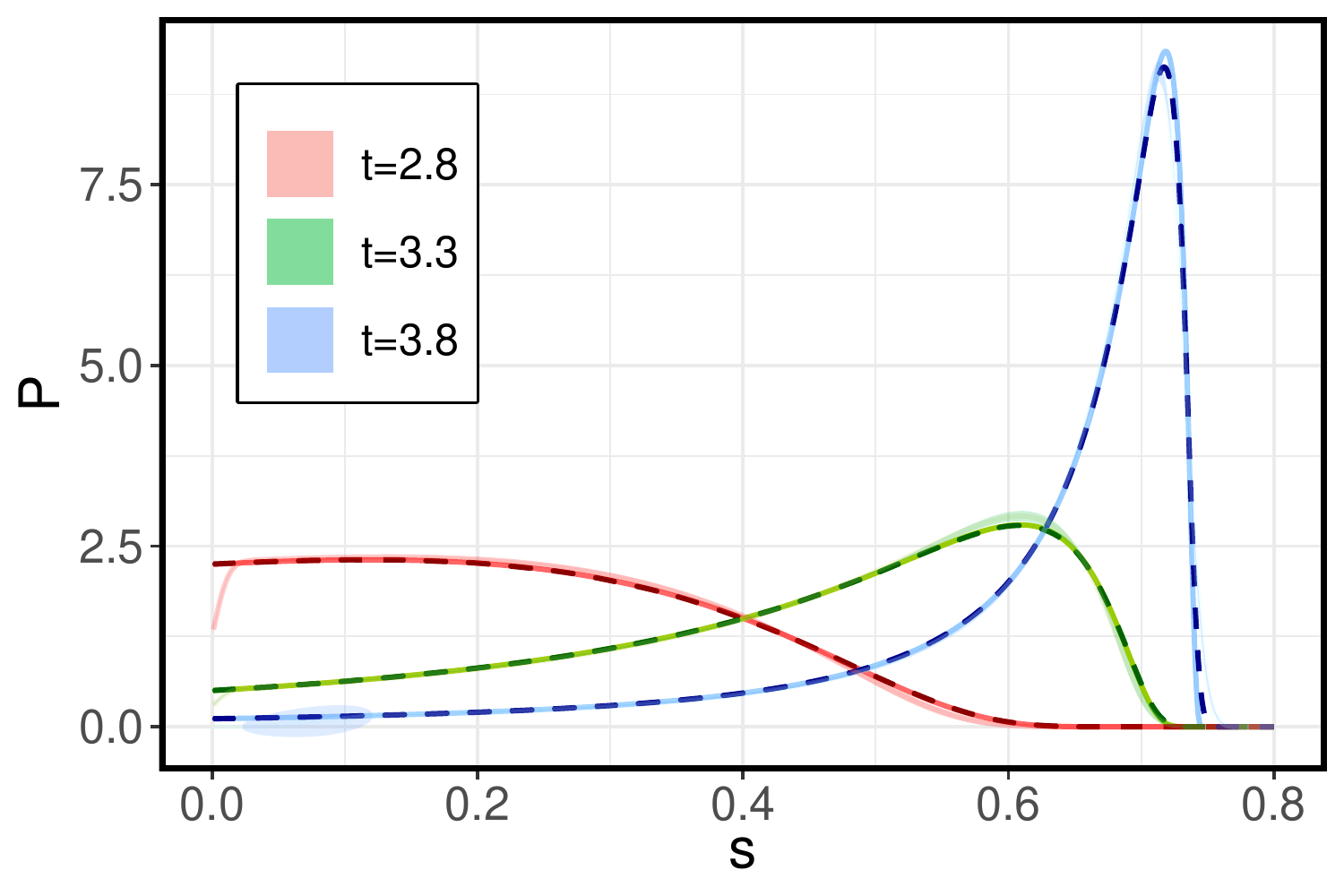} 
    \caption{The operator size distribution of the Brownian circuits is shown for $N=10^4$. In this plot, different colors correspond to different times $t$. The dashed lines represent the results obtained from numerical simulations, while the solid lines depict the analytical solution given by Eq. \eqref{eq:Brownianres}. The region shaded with light color is obtained by numerically evaluating the integral in \eqref{eq:relation} for various $t_0$ values, while ensuring that $\overline{s_0}/s_{\text{sc}}\in [0.03,0.05]$.
    %Comparison of size distribution of the numerical result using Eq. \eqref{eq:relation}, the analytical result Eq. \eqref{eq:Brownianres} and the master equation. The large error at small size comes from the approxiamation of delta function.
    }
\end{figure}

\textbf{Protocol}: We summarize the concrete experimental protocol to test the validity of the SEFT on quantum simulators:

\textit{Step 1}: Perform measurements of the operator size distribution $\mathcal{P}(s,t)_{\text{exp}}$ for different times $t$ with $\hat{O}=\hat\sigma^j_\alpha$, following the size-total spin relation or other existing protocols \cite{Qi:2019rpi,Blocher:2023hvk}. Compute the averaged distribution $\overline{\mathcal{P}}(s,t)_{\text{exp}}$ over sites $j$ and $\alpha\in\{x,y,z\}$.

\textit{Step 2}: Choose a set of ${t_0^k}$ with $t_0^k \ll \log N$, compute the averaged operator size $\overline{s_0}$, and extract $\varkappa$ by performing a linear fit of the averaged size $\log \overline{s_0}$ for different $t_0$ values.

\textit{Step 3}: Select a specific $t_0 \ll \log N$, compute the numerical integral in Eq. \eqref{eq:relation} using the experimental result $\overline{\mathcal{P}}(s,t_0)_{\text{exp}}$ for a general time $t$ to obtain the theoretical prediction $\overline{\mathcal{P}}(s,t)_{\text{th}}$. Compare this prediction with the experimental data $\overline{\mathcal{P}}(s,t)_{\text{exp}}$.

\vspace{5pt}

We finally address the generalization to systems without time-reversal symmetry or at finite temperatures. (i). In the absence of time-reversal symmetry, the vertices for absorbing scramblons in the future and emitting scramblons in the past become distinct, denoted as $\Upsilon^{\text{R},m}$ and $\Upsilon^{\text{A},m}$, respectively. Similarly, the labels R$/$A should be added to the functions $f_O^{\text{R/A}}$ and $h_O^{\text{R/A}}$. Consequently, we can only determine the operator size at late-time $t\gg t_{\text{sc}}$ by combining information of $\mathcal{P}(s,t_0)$ and $\mathcal{P}(s,-t_0)$. (ii). At finite temperatures, an unambiguous definition of the operator size distribution is lacking. To establish a relationship between the operator sizes in different time regimes, we adopt a specific definition. The operator size distribution of $\hat{O}(t)$ at an inverse temperature $\beta$ is defined as the standard operator size distribution of $\rho^{1/4}\hat{O}(t)\rho^{1/4}$, where $\rho=e^{-\beta \hat{H}}$ represents the thermal density matrix. Under this definition, an analog of \eqref{eq:relation} holds. We leave a detailed derivation into the supplementary material \cite{SM}. 

\vspace{5pt}

     \emph{ \color{blue!60}Example.--}    
As an illustration, let us consider the Brownian circuits for spin-$1/2$ systems \cite{Zhou:2018snw}. The Hamiltonian reads
\begin{equation}
{H_s}\Delta t = J\sum_{i < j} \sum_{{\mu _i},{\mu _j} } \hat{\sigma} ^i_{{\mu_i}}  \hat{\sigma} ^j_{{\mu_j}}\Delta B^{i,j}_{{\mu _i},{\mu _j}},
\end{equation}
where $\mu_i\in \{0,x,y,z\}$. $\Delta B^{i,j}_{{\mu _i},{\mu _j}}$ represents independent Brownian variables with $(\Delta B^{i,j}_{{\mu _i},{\mu _j}})^2=\Delta t$. We take the normalization $J = {({{{8}N}})^{-1/2}}$. The system exihibits the time-reversal symmetry. In Ref. \cite{Zhou:2018snw}, the authors solved the operator size distribution of Brownian circuits by deriving a set of ordinary equations for $P(n,t)$ using It\^o Calculus, which allows for efficient classical simulations with large $N$. However, due to the lack of a diagrammatic analysis, it remains unclear whether information scrambling in Brownian circuits can be described by the SEFT.

Here, we apply our protocol to Brownian circuits to address this question. In Brownian circuits, there is a permutation symmetry for different spins $j$ and different directions $\alpha$. As a result, the average can be omitted simplifying the analysis. The early-time operator size distribution $P(n,t)$ has been analytically computed in \cite{Zhou:2018snw}. Taking the thermodynamic limit with fixed $e^{\varkappa t}/N$ and $\varkappa=3$, we discover
\begin{equation}
\begin{aligned}\label{eq:Brownianearly}
\overline{\mathcal{P}}(s,t_{0}) =Ne^{- \varkappa t_{0}}\text{exp}\big(-sNe^{- \varkappa t_{0}}\big),\ \ \ \ \ \ \overline{s_0}=e^{ \varkappa t_{0}}/N.
\end{aligned}
\end{equation}
We then try to predict the operator size distribution for general time $t$ using \eqref{eq:relation} and \eqref{eq:Brownianearly}. Performing the integration gives
\begin{equation}\label{eq:Brownianres}
\overline{\mathcal{P}}(s,t)=\frac{N}{(1-s/s_{\text{sc}})^{2}}\exp(-\varkappa t-\frac{sNe^{-\varkappa t}}{(1-s/s_{\text{sc}})}).
\end{equation}
In FIG. 3, we compare this result to numerical simulations with $N={10^4}$. We find excellent agreement for arbitrary time $t$. This supports the validity of the SEFT in Brownian circuits. Interestingly, this formula has not been obtained explicitly within to our knowledge. Its moments match Eq. (20) in \cite{Zhou:2018snw} . For example, we have $\overline{s}/s_{\text{sc}}=1+a e^a \text{Ei}(-a)$ with $a=s_{\text{sc}}Ne^{-\varkappa t}$. For smaller $N$, we should take finite size corrections for Eq. \eqref{eq:relation} into account, as established in the supplementary material \cite{SM}.
\vspace{5pt}

Up to now, we are using the analytical expression \eqref{eq:Brownianearly} for $\mathcal{P}(s,t_0)$. However, the extraction of $\mathcal{P}(s,t_0)$ contains additional continuous limit, which may lead to large error. We further test our protocol by direct application to numerical data of $P(n,t)$. We choose $t_0$ by fixing $\overline{s_0}/s_{\text{sc}}\in [0.03,0.05]$. The integral in \eqref{eq:relation} is performed numerically by using a Gaussian approximation of the Dirac delta function with standard deviation $\sigma={10^{-2}}$. The result for different $t_0$ is plotted in FIG. 3 as the shaded region, which is consistent with both numerics and analytical results \eqref{eq:Brownianres}.

\vspace{5pt}
\emph{ \color{blue!60}MQC.--} 
After confirming the validity of the SEFT in a specific model, we can proceed to make theoretical predictions for other experimental observables. As an example, we present results for the MQC, which is extensively studied in NMR experiments \cite{abraham1998introduction}. To obtain MQC of $\hat{O}$, one first performs the measurement of
\begin{equation}
F(\varphi,t)=\langle e^{i\varphi\sum_j\hat{s}_z^j }\hat{O}(t)e^{-i\varphi\sum_j\hat{s}_z^j }\hat{O}(t)\rangle,\ \ \ \ \ \ \ \varphi\in[0,2\pi).
\end{equation}
The MQC, denoted as $I(m,t)$, is defined as the Fourier transform of $F(\varphi,t)$ for integer values of $m$. In the thermodynamic limit $N\gg 1$, $F(\varphi,t)$ rapidly concentrates around $\varphi=0$ with a width on the order of $\sim{N}^{-\frac{1}{2}}$. To address this, we introduce the variable $\xi\equiv \sqrt{N}\varphi$ and $\mathcal{F}(\xi,t)\equiv F(\sqrt{N}\phi,t)$. 
\begin{comment}
Similar to the generating function for operator sizes, we can demonstrate that
\begin{equation}
\mathcal{F}(\xi,t)=\langle \hat{O}(t)|e^{-i\frac{\xi}{2\sqrt{N}}\sum_j{(\sigma}^j_z+ {\tau}^j_z)}|\hat{O}(t)\rangle.
\end{equation}
\end{comment}
The computation of $\mathcal{F}(\xi,t)$ can be done in a similar manner as for $\mathcal{S}(\nu,t)$ utilizing the SEFT. The resulting expression is 
\begin{equation}
\mathcal{F}(\xi,t)=\int_0^\infty dy~h_O(y) e^{-\frac{\xi^2}{4}(1-f_{z}(\lambda y))}.
\end{equation}
Unlike the operator size distribution, the MQC solely depends on $f_z$. By performing the Fourier transform $\mathcal{I}(u,t)=\int d\xi e^{i u \xi} \mathcal{F}(\xi,t)/2\pi$, the continuum limit of the MQC, denoted as $\mathcal{I}(u,t)$ with $u\equiv m/\sqrt{N}$, can be expressed as
\begin{equation}
\mathcal{I}(u,t)=\int_0^\infty dy~h_O(y)[\pi(1-f_z(\lambda y))]^{-1/2}e^{-\frac{u^2}{1-f_z(\lambda y)}}.
\end{equation}
This can also be related to the operator size distribution at early times. For instance, if we consider $O=\hat{\sigma}^j_z$ and take an average over sites $j$, we can show that
\begin{equation}\notag
\mathcal{I}_z(u,t)=\int_0^\infty ds_1~\mathcal{P}_z(s_1,t_0)[\pi(1-\mathcal{S}_z(\eta s_1,t_0))]^{-1/2}e^{-\frac{u^2}{1-\mathcal{S}_z(\eta s_1,t_0)}}, 
\end{equation}
where we have introduced $\eta(t,t_0)={e^{\varkappa(t-t_{0})}}/{s_{\text{sc}}\overline{s_{0}}}$ for conciseness. The subscripts $z$ are included as a reminder.

\vspace{5pt}
\emph{ \color{blue!60}Discussions.--}  
In this work, we explore the signature of the SEFT in random spin models with all-to-all interactions. Our result \eqref{eq:relation} reveals a non-trivial consistency relation between the distribution of operator sizes in the early-time regime and their values at late-time. This finding serves as strong evidence for the validity of SEFT in a specific model, which can be experimentally tested on quantum simulator platforms. As an illustration, we apply the scheme to Brownian circuits, where a diagrammatic calculation is not available. Both analytical and numerical implements of our protocol confirms the validity of SEFT in Brownian circuits. We also present discussions for the MQC. 

Recently, there are growing interest in quantum teleportation through emergent traversable wormholes \cite{Gao:2016bin,Maldacena:2017axo,Susskind:2017nto,Gao:2018yzk,Brown:2019hmk,Gao:2019nyj,Nezami:2021yaq,PhysRevX.12.031013}, which has been experimentally investigated in Ref. \cite{Jafferis:2022crx}. The teleportation fidelity has been proposed to be associated with the size winding at finite temperatures. Our calculation can be generalized to establish a relationship between size winding in different time regimes for systems described by the SEFT. We would like to postpone a detailed study to future works.

\vspace{5pt}

\emph{Acknowledgement.} 
We thank Xiao Chen, Yingfei Gu, Xinhua Peng and Tian-Gang Zhou for invaluable discussions.

\vspace{5pt}

%\textit{Acknowledgment.} We thank Xiao Chen and Yingfei Gu for helpful discussions. 

\bibliography{draft__0531.bbl}

\ifarXiv


\section*{Supplementary material}

# Supplemental Material for "Signature of Scramblon Effective Field Theory in Random Spin Models"

Zeyu Liu[1] and Pengfei Zhang[1,2,*]

[1]*Department of Physics, Fudan University, Shanghai, 200438, China*
[2]*Center for Field Theory and Particle Physics, Fudan University, Shanghai, 200438, China*
(Dated: June 9, 2023)

In this supplemental material, we present: (i). Detailed derivation of operator size without time-reversal symmetry or at finite temperature; (ii). Numerics for the Brownian circuit model; (iii). Finite size correction.

## I. SYSTEMS WITHOUT TIME-REVERSAL SYMMETRY OR AT FINITE TEMPERATURE.

At finite temperatures, an unambiguous definition of the operator size distribution is lacking. To establish a relationship between the operator sizes in different time regimes, we adopt a specific definition. The operator size distribution of $O(t)$ at an inverse temperature $\beta$ is defined as the standard operator size distribution of $\rho^{1/4} O(t) \rho^{1/4}$, where $\rho = e^{-\beta H}$ represents the thermal density matrix. We start with continuous size distribution

$$\mathcal{P}(s,t) = \langle \rho^{1/4} O(t) \rho^{1/4} | \delta(s - \sum_j (s^j_{\text{tot}})^2/2N) | \rho^{1/4} O(t) \rho^{1/4} \rangle. \tag{1}$$

Then the generating function $\mathcal{S}(\nu,t) = \int ds\, \mathcal{P}(s,t) e^{-\nu s}$ is given by

$$\mathcal{S}(\nu,t) = e^{-\frac{3\nu}{4}} \langle \rho^{1/4} O(t) \rho^{1/4} | e^{-\frac{\nu}{4N} \sum_j \sigma^j \cdot \tau^j} | \rho^{1/4} O(t) \rho^{1/4} \rangle. \tag{2}$$

For conciseness, we introduce the complex time $\theta = \tau + it$, and we represent (2) in the path integral presentation as

$$\mathcal{S}(\nu,t) = e^{-\frac{3\nu}{4}} \langle T_c\, O(\theta_1) O(\theta_2) e^{-\frac{\nu}{4N} \sum_j \sigma^j(\theta_3) \cdot \tau^j(\theta_4)} \rangle. \tag{3}$$

We have introduced

$$\theta_1 = it, \qquad \theta_2 = -\frac{\beta}{2} + it, \qquad \theta_3 = -\frac{\beta}{4}, \qquad \theta_4 = -\frac{3\beta}{4}. \tag{4}$$

Assuming the validity of the scramblon description, the generating function $\mathcal{S}(\nu,t)$ can be computed in closed-form in the large N limit [1–4]. Performing the Taylor expansion of (2), the result contains general OTOCs between $O$ and Pauli strings

$$\mathcal{S}(\nu,t) = e^{-\frac{3\nu}{4}} \sum_n \left(\frac{\nu}{4N}\right)^n \sum_{\{j_k,\alpha_k\}} \langle O(\theta_1) \sigma^{j_1}_{\alpha_1}(\theta_3)...\sigma^{j_n}_{\alpha_n}(\theta_3) O(\theta_2) \sigma^{j_1}_{\alpha_1}(\theta_4)...\sigma^{j_n}_{\alpha_n}(\theta_4) \rangle, \tag{5}$$

where we have used the property $s^j_{\text{tot}}|\text{EPR}\rangle = (s'^j + s^j)|\text{EPR}\rangle = 0$. Summing up all diagrams with arbitrary number of scrambling modes and neglect contributions from collision terms, we have

$$\mathcal{S}(\nu,t) = \vcenter{\hbox{[diagram]}} = e^{\frac{-3\nu}{4}} \sum_n \frac{1}{n!} \left(\frac{\nu}{4}\right)^n \sum_{\{m_k,\alpha_k\}} \frac{(-\lambda)^{\sum_{l=1}^n m_l}}{m_1! \cdots m_n!} \Upsilon_O^{R,\sum_{l=1}^n m_l}(\theta_1 - \theta_2) \Upsilon_{\alpha_1}^{A,m_1}(\theta_3 - \theta_4) \cdots \Upsilon_{\alpha_n}^{A,m_n}(\theta_3 - \theta_4), \tag{6}$$

---

[*] PengfeiZhang.physics@gmail.com

where $\lambda = C^{-1} e^{i\varkappa(\beta/2 - \theta_1 - \theta_2 + \theta_3 + \theta_4)/2} = C^{-1} e^{\varkappa t}$. For convenience, we introduce functions

$$f_\alpha^{R/A}(x,\theta) = \sum_m \frac{(-x)^m}{m!} \Upsilon_\alpha^{R/A,m}(\theta) = \int_0^\infty dy\, h_\alpha^{R/A}(y,\theta) e^{-xy}, \qquad \Upsilon_\alpha^{R/A,m}(\theta) = \int_0^\infty dy\, h_\alpha^{R/A}(y,\theta) y^m. \qquad (7)$$

It is straight forward to show that

$$\overline{S}(\nu,t) = \int_0^\infty dy\, \overline{h}^R(y,\beta/2) \exp\left(-\nu s_{\text{sc}}(1 - \overline{f}^A(\lambda y, \beta/2))\right), \qquad (8)$$

with $s_{\text{sc}} = \frac{3}{4}$ as the typical size for a maximally scrambled operator in spin model, and $\overline{g}(x) \equiv \sum_{\alpha,j} g_{\sigma_\alpha^j}(x)/3N$. By the inverse Laplace transform, we can obtain the size distribution

$$\overline{\mathcal{P}}(s,t) = \int_0^\infty dy\, \overline{h}^R(y,\beta/2) \delta\left(s - s_{\text{sc}} + s_{\text{sc}} \overline{f}^A(\lambda y,\beta/2)\right). \qquad (9)$$

We consider the following three conditions:
(1) Infinite temperature with time-reversal symmetry:

$$\overline{S}(\nu,t) = \int_0^\infty dy\, \overline{h}(y,0) \exp\left(-\nu s_{\text{sc}}(1 - \overline{f}(\lambda y, 0))\right), \qquad (10)$$

which is exactly what we get in the main text, the key result is

$$\overline{\mathcal{P}}(s,t) = \int_0^\infty ds_1\, \overline{\mathcal{P}}(s_1,t_0) \delta\left(s - s_{\text{sc}} + s_{\text{sc}} \int_0^\infty ds_2\, \overline{\mathcal{P}}(s_2,t_0) \exp\left(-\frac{s_1 s_2}{s_{\text{sc}} \overline{s}_0} e^{\varkappa(t-t_0)}\right)\right). \qquad (11)$$

(2) Infinite temperature without time-reversal symmetry:

Similar to what we do in main text, in the early-time regime with $N^{-1} \ll \lambda_0 = e^{\varkappa t_0}/C \ll 1$, we expand $\overline{f}^A(\lambda_0 y) = 1 - \lambda_0 y \overline{\Upsilon}^{A,1}$ to get

$$\overline{\mathcal{P}}(s,t_0) = \frac{\overline{\Upsilon}^{R,1}}{\overline{s}_0} \overline{h}^R(s \overline{\Upsilon}^{R,1}/\overline{s}_0), \qquad \overline{\mathcal{P}}(s,-t_0) = \frac{\overline{\Upsilon}^{A,1}}{\overline{s}_0} \overline{h}^A(s \overline{\Upsilon}^{A,1}/\overline{s}_0), \qquad \overline{s}_0 = \int_0^1 ds\, s \overline{\mathcal{P}}(s,t_0) = s_{\text{sc}} \lambda_0 \overline{\Upsilon}^{A,1} \overline{\Upsilon}^{R,1}. \qquad (12)$$

Note that $\overline{s}_0$ is time-reversal-invariant, and the generating function can be expressed as

$$\overline{S}(\nu,t_0) = \overline{f}^R(\nu \overline{s}_0 / \overline{\Upsilon}^{R,1}), \qquad \overline{S}(\nu,-t_0) = \overline{f}^A(\nu \overline{s}_0 / \overline{\Upsilon}^{A,1}). \qquad (13)$$

Using (9)(12)(13), we find

$$\overline{\mathcal{P}}(s,t) = \int_0^\infty ds_1\, \overline{\mathcal{P}}(s_1,t_0) \delta\left(s - s_{\text{sc}} + s_{\text{sc}} \int_0^\infty ds_2\, \overline{\mathcal{P}}(s_2,-t_0) \exp\left(-\frac{s_1 s_2}{s_{\text{sc}} \overline{s}_0} e^{\varkappa(t-t_0)}\right)\right). \qquad (14)$$

(3) Finite temperature with time-reversal symmetry:

In the case of finite temperature, a key point is $G(\beta/2) \equiv \overline{\Upsilon}^0(\beta/2) \neq 1$. In the early-time regime, we expand $\overline{f}(\lambda_0 y, \beta/2) = G(\beta/2) - \lambda_0 y \overline{\Upsilon}^1(\beta/2)$, and further get

$$\overline{\mathcal{P}}(s,t_0) = (s_{\text{sc}} \lambda_0 \overline{\Upsilon}^1(\beta/2))^{-1} \overline{h}\left((s - s_{\text{th}}) s_{\text{sc}} \lambda_0 \overline{\Upsilon}^1(\beta/2)^{-1}\right), \qquad \overline{s}_0 = s_{\text{th}} G(\beta/2) + s_{\text{sc}} \lambda_0 \overline{\Upsilon}^1(\beta/2)^2, \qquad (15)$$

where $s_{\text{th}} = s_{\text{sc}}(1 - G(\beta/2))$ is the typical operator size of the thermal density matrix $\rho^{1/2}$, similarly we have

$$\overline{\mathcal{P}}(s,t) = \int_{s_{\text{th}}}^\infty ds_1\, \overline{\mathcal{P}}(s_1,t_0) \delta\left(s - s_{\text{sc}} + s_{\text{sc}} \int_{s_{\text{th}}}^\infty ds_2 \overline{\mathcal{P}}(s_2,t_0) \exp\left(-\frac{(s_1 - s_{\text{th}})(s_2 - s_{\text{th}})}{s_{\text{sc}}(\overline{s}_0 - G(\beta/2) s_{\text{th}})} e^{\varkappa(t-t_0)}\right)\right). \qquad (16)$$

It coincide with (11) when $\beta = 0$.

## II. NUMERICS FOR THE BROWNIAN CIRCUIT MODEL

We apply the method developed in Ref. [5] to our numerics for the Brownian circuit model presented in Fig. 3 in the main text. The master equation of operator size $P(n,t)$ is given by

$$\frac{dP(t)}{dt} = A_P P(t), \tag{17}$$

with a tri-diagonal stochastic matrix $A_P$

$$\begin{aligned}
(A_P)_{k,k} &= \frac{4}{N}k[-(N-k) + \frac{1}{4}(N-2k+1)], \\
(A_P)_{k-1,k} &= \frac{1}{N}k(k-1), \\
(A_P)_{k+1,k} &= \frac{3}{N}k(N-k).
\end{aligned} \tag{18}$$

And we have an analytical result of $\overline{\mathcal{P}}(s,t)$ given by Eq. (16) in the main text

$$\overline{\mathcal{P}}(s,t) = \frac{N}{(1-s/s_{sc})^2} \exp\left(-\varkappa t - \frac{sNe^{-\varkappa t}}{(1-s/s_{sc})}\right). \tag{19}$$

We test our protocol using (11)

$$\overline{\mathcal{P}}(s,t) = \int_0^\infty ds_1 \, \overline{\mathcal{P}}(s_1,t_0) \delta\left(s - s_{sc} + s_{sc} \int_0^\infty ds_2 \, \overline{\mathcal{P}}(s_2,t_0) \exp\left(-\frac{s_1 s_2}{s_{sc} \overline{s_0}} e^{\varkappa(t-t_0)}\right)\right). \tag{20}$$

We choose $t_0$ by fixing $\overline{s_0}/s_{sc} \in [0.03, 0.05]$. The integral in is performed numerically by using a Gaussian approximation of the Dirac delta function with standard deviation $\sigma = 10^{-2}$, which reads $g(x) = \frac{1}{\sqrt{2\pi\sigma^2}} \exp\left(-\frac{x^2}{2\sigma^2}\right)$. The result for different $t_0$ is plotted in FIG. 3 in the main text as the shaded region, which is consistent with both numerics (17) and analytical results (19).

## III. FINITE N CORRECTION

We start with the generating function at infinite temperature

$$\mathcal{S}(\nu,t) = e^{-\frac{3\nu}{4}} \langle O(t)|e^{-\frac{\nu}{4N}\sum_j \sigma^j \cdot \tau^j}|O(t)\rangle. \tag{21}$$

We strictly rewrite the generating function in the following term

$$\begin{aligned}
\mathcal{S}(\nu,t) &= e^{-\frac{3\nu}{4}} \langle O(t)| \prod_{j=1}^N \prod_{\alpha=1}^3 \left(\cosh(\nu/4N) - \sinh(\nu/4N)\sigma_\alpha^j \tau_\alpha^j\right)|O(t)\rangle \\
&= e^{-\frac{3\nu}{4}} \langle O(t)| \prod_{j=1}^N \left(\cosh^3(\nu/4N) + \sinh^3(\nu/4N) - (\cosh(\nu/4N) + \sinh(\nu/4N))\sinh(\nu/4N)\cosh(\nu/4N)\sigma_\alpha^j \tau_\alpha^j\right)|O(t)\rangle.
\end{aligned} \tag{22}$$

Summing up all diagrams with arbitrary number of scrambling modes and neglect contributions from collision terms like (6), we get

$$\mathcal{S}(\nu,t) = e^{-\frac{3\nu}{4}} \int_0^\infty dy \, h_O^R(y) \prod_{j=1}^N \left((\frac{1}{4}e^{3\nu/4N} + \frac{3}{4}e^{-\nu/4N}) + \frac{1}{4}\sum_{\alpha=1}^3 f_\alpha^{j,A}(\lambda y)(e^{3\nu/4N} - e^{-\nu/4N})\right). \tag{23}$$

For chaotic models with $\overline{f}(\infty) = 0$ [2], when $\lambda \to \infty$, we have

$$\mathcal{S}(\nu,\infty) = \prod_{j=1}^N (\frac{1}{4} + \frac{3}{4}e^{-\nu/N}). \tag{24}$$

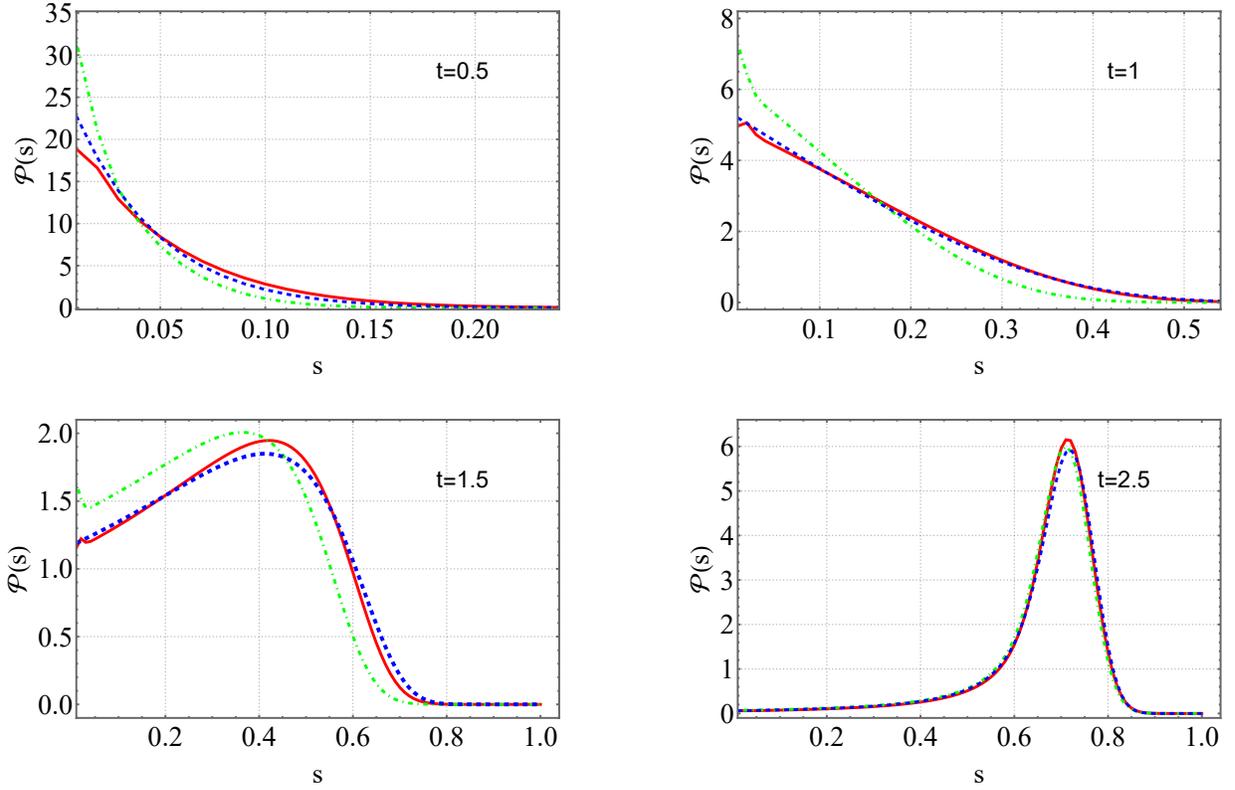

FIG. 1. Operator size distribution $\overline{\mathcal{P}}(s,t)$ at four different times with $N = 10^2$. Different graphs correspond to different times $t$. The blue dashed lines represent the results obtained from (17), while the red solid lines depict the analytical solution given by Eq. (30) using $\overline{\mathcal{P}}(s,t_0) = Ne^{-\varkappa t_0}\exp(-sNe^{-\varkappa t_0})$. The green dotted lines are obtained by numerically evaluating the integral in (30) with $t_0 = 0.7$.

Operator size distribution is given by

$$P(n,\infty) = \binom{N}{n}(s_{\rm sc})^n(1-s_{\rm sc})^{N-n}, \tag{25}$$

which is the Binomial distribution. Further we can regard $\nu/4N$ as small parameter and expand (23) to order of $\nu/4N$ in the exponent and it reads

$$\overline{S}(\nu,t) = \int_0^\infty dy\, \overline{h}^R(y) \exp\left(-\nu s_{\rm sc}(1-\overline{f}^A(\lambda y))\right), \tag{26}$$

which coincide with (8). Expanding (23) to order of $(\nu/4N)^2$ leads to

$$\overline{S}(\nu,t) = \int_0^\infty dy\, \overline{h}^R(y,\beta/2) \exp\left(-s_{\rm sc}\nu(1-\overline{f}^A(\lambda y)) + \frac{3\nu^2}{32N}(1+2\overline{f}^A(\lambda y)-3\overline{f^2}^A(\lambda y))\right), \tag{27}$$

where $\overline{f^2}^A(\lambda y) \equiv \frac{1}{N}\sum_j(\frac{1}{3}\sum_\alpha f_\alpha^{j,A}(\lambda y))^2$. By the inverse Laplace transform, we can obtain the size distribution

$$\overline{\mathcal{P}}(s,t) = \int_0^\infty dy\, \overline{h}^R(y) \frac{1}{\sqrt{2\pi\sigma^2}} \exp\left(-\frac{1}{2\sigma^2}(s-s_{\rm sc}(1-\overline{f}^A(\lambda y)))^2\right), \tag{28}$$

where

$$\sigma^2 = \frac{s_{\rm sc}(1-s_{\rm sc})\left(1+2\overline{f}^A(\lambda y)-3\overline{f^2}^A(\lambda y)\right)}{N}. \tag{29}$$

The difference between large-N solution (11) is that the delta function is replaced by a finite width Gaussian function centered at the same position. Assuming time-reversal-symmetry, we have

$$\overline{\mathcal{P}}(s,t) = \int_0^\infty ds_1\, \overline{\mathcal{P}}(s_1,t_0) \frac{1}{\sqrt{2\pi\sigma^2}} \exp\Big(-\frac{1}{2\sigma^2}(s - s_{\text{sc}}(1 - \overline{\mathcal{S}}(\eta s_1, t_0))^2\Big). \tag{30}$$

$$\sigma^2 = \frac{s_{\text{sc}}(1 - s_{\text{sc}})\big(1 + 2\overline{\mathcal{S}}(\eta s_1, t_0) - 3\overline{\mathcal{S}}(\eta s_1, t_0)^2\big)}{N}. \tag{31}$$

We have introduced $\eta(t, t_0) = e^{\varkappa(t-t_0)}/s_{\text{sc}}\overline{s_0}$ for conciseness. $\overline{\mathcal{S}}(\eta s_1, t_0) = \int_0^\infty ds_2\, \overline{\mathcal{P}}(s_2, t_0) \exp(-\eta s_1 s_2)$ is the generating function and $\overline{s_0} = \int_0^\infty ds_2\, \overline{\mathcal{P}}(s_2, t_0) s_2$ is the operator size. The integral in (30) is performed numerically, the result for different $t$ is plotted in FIG. 1, which is consistent with (17).

---


\fi
	
\end{document}